\documentclass[slac_one]{revtex4}
\usepackage{graphicx}
\usepackage{fancyhdr}
\usepackage{amsmath}
\usepackage{amssymb}
\newcommand{\abs}[1]{\left\lvert #1\right\rvert}

\newcommand{\vect}[1]{\mathbf{#1}}
\newcommand{\bra}[1]{\left\langle #1\right\rvert}
\newcommand{\ket}[1]{\left\lvert #1\right\rangle}
\newcommand{\Lqcd}{\Lambda_{\text{QCD}}}
\newcommand{\bea}{\begin{eqnarray}}
\newcommand{\eea}{\end{eqnarray}}
\newcommand{\e}{\mathrm{e}}

\DeclareMathOperator{\Tr}{Tr}

\begin{document}

\title{Universality of Nonperturbative Effects in Event Shapes}


\author{Christopher Lee}
\affiliation{Institute for Nuclear Theory, University of Washington, Box 351550, Seattle, WA  98195-1550 USA}
\author{George Sterman}
%%%MY ADDRESS
\affiliation{C.N.\ Yang Institute for Theoretical Physics, Stony Brook University,
Stony Brook, NY 11794-3840 USA }

\begin{abstract}
Nonperturbative effects in event shape distributions can be characterized by shape functions derived in the eikonal approximation or, equivalently, from soft-collinear effective theory. The use of energy flow operators and the boost invariance of the Wilson lines of soft gluons in the shape functions leads to a proof of universality for power corrections to the mean values of event shapes, without invoking the single gluon approximation.

\end{abstract}

\maketitle

\thispagestyle{fancy}


In this Comment, we elaborate  the 
 formalism of soft-collinear effective theory (SCET)
presented in \cite{T003} for the shape functions for thrust, $C$-parameter, and angularities.
We show that  several well-known universality relations for mean values of these observables
can be derived
without invoking the approximation of single soft gluon emission. 
These relations
can be derived either in the language of SCET at leading order in the expansion parameter $\lambda=\sqrt{\Lqcd/Q}$ \cite{BMW,BLMW} or, equivalently, in the eikonal approximation \cite{KS95,KS99}.

The central result described in \cite{T003} 
is an operator expression for
the nonperturbative shape function for the distribution in the event shape $e$ (where $e\rightarrow 0$ in the two-jet limit):
\begin{equation}
\label{shapefunction}
S_e(e) = \frac{1}{\sigma_0}\frac{d\sigma}{de} = \frac{1}{N_C}\Tr\sum_{X_u}\abs{\bra{X_u}Y_n\overline Y_{\bar n}\ket{0}}^2\delta(e- e(X_u)),
\end{equation}
where $\sigma_0$ is the total cross-section at leading-order in $\alpha_s$, $N_C$ is the number of colors, the sum is over final states $X_u$ in the ultrasoft sector, and $Y_n, \overline Y_{\bar n}$ are Wilson lines of ultrasoft gluons corresponding the outgoing quark and antiquark in the $n,\bar n$ light-cone directions. The delta function fixes the contribution to the event shape 
$e(X_u)$ for each final state $X_u$ in the ultrasoft sector 
equal to a value $e$.   We will develop below
a way to eliminate the explicit sum over these states.

The event shapes we will consider are the $C$-parameter and angularities (which include the thrust). It will be useful to re-express these event shapes in terms of the pseudorapidities $\eta_i$ of the final state particles, where
\begin{equation}
\eta_i = \ln\cot\frac{\theta_i}{2},
\end{equation}
measured with respect to the thrust axis $\vect{\hat t}$. For massless particles, these are equivalent to the rapidities
\begin{equation}
\vartheta_i = \frac{1}{2}\ln\frac{E_i + p^z_i}{E_i - p^z_i},
\end{equation}
with the $z$-axis taken to be along $\vect{\hat t}$. In terms of the pseudorapidities, the angularities \cite{scaling1} and $C$-parameter \cite{Salam01} can be expressed as:
\begin{align}
\tau_a &= \frac{1}{Q}\sum_{i\in N}\abs{\vect{p}_i^\perp} \e^{-\abs{\eta_i}(1-a)}\, , \\
C &= \frac{1}{Q}\sum_{i\in N}\frac{3\abs{\vect{p}_i^\perp}}{\cosh\eta_i}.
\end{align}
In this form, the relations between the event shapes 
implied by invariance
under Lorentz boosts will be made more manifest.

A striking  prediction from the analysis of these event shapes \cite{KS95}-\cite{flow} 
is the universality in power corrections to their mean values,
\bea
\langle e \rangle =  \langle e \rangle_{\text{PT}} + c_e \mathcal{A}\, ,
\label{avgshift}
\eea
with ${\cal A}$ a universal parameter and $c_e$ a
calculable coefficient that depends on the observable,
\begin{equation}
\label{Cfactors}
c_{\tau_a} = \frac{2}{1-a},\qquad c_{C} = 3\pi. 
\end{equation}
(We limit our attention to angularities for $a<1$.) 
In an approximation that we discuss below, 
their effect also is to shift the event shape distributions,
\begin{equation}
\label{shift}
\frac{d\sigma}{d e}(e)\biggr\rvert_{\text{PT}} \underset{\text{NP}}{\longrightarrow} \frac{d\sigma}{de}(e - c_e \mathcal{A})\biggr\rvert_{\text{PT}} .
\end{equation}
These relations were derived in  Refs. \cite{DW1,DMW,DW2} from 
the assumption of a ``dispersive" representation for
$\alpha_s(\mu^2)$ considered as an analytic function
of the scale $\mu$.  
In Refs.\  \cite{KS95},
  they were abstracted directly from the form of resummed perturbation
theory.
A more general approach in \cite{KS99,scaling1,scaling2} replaces the shift by a convolution with
the shape function defined as above.  In \cite{KS99} and \cite{flow} the role of energy flow 
was explored in a manner closely related to our discussion below.

While we have seen how the single gluon approximation may be implemented in the effective theory \cite{T003} to reproduce the well-known relations between the shifts to event shape distributions, the shape function (\ref{shapefunction}) already contains within itself all the information needed to prove Eq.\ (\ref{avgshift}) for the mean values,
 without making any further approximations. We simply need a few tools to unveil this information.  We will concentrate here on the shape function associated with ultrasoft
radiation.  Its factorization from perturbative collinear radiation is discussed in the context of SCET
in \cite{T003,BMW,BLMW}, or equivalently in \cite{KS99}.

Consider first the angularity distribution. In this case, the shape function is
\begin{equation}
S_{\tau_a}(\tau_a) = \frac{1}{N_C} \Tr\sum_{X_u}\bra{0}\overline{Y}_{\bar n}^\dag Y_n^\dag \ket{X_u}\bra{X_u} Y_n \overline{Y}_{\bar n}\ket{0}\delta\left(\tau_a - \frac{1}{Q}\sum_{i\in X_u}\abs{\vect{k}_i^\perp} \e^{-\abs{\eta_i}(1-a)}\right).
\end{equation}
We now define a transverse energy flow operator, by analogy to the energy flow operator introduced in Ref.~\cite{flow}, defined through its action on a state $N$:
\begin{equation}
\mathcal{E}_T(\eta)\ket{N} = \sum_{i\in N}\abs{\vect{k}_i^\perp}\delta(\eta - \eta_i)\ket{N},
\end{equation}
where the sum is over the particles $i$ in state $N$.
Making use of this operator, we can write the shape function as
\begin{equation}
S_{\tau_a}(\tau_a) = \frac{1}{N_C}\Tr\sum_{X_u}\bra 0 \overline{Y}_{\bar n}^\dag Y_n^\dag \delta\left(\tau_a - \frac{1}{Q}\int\! d\eta\,\e^{-\abs{\eta}(1-a)}\mathcal{E}_T(\eta)\right)\ket{X_u}\bra{X_u}Y_n\overline{Y}_{\bar n}\ket{0}.
\end{equation}
The operators in the matrix element no longer contain any reference to the final state $X_u$, so we may perform the sum over intermediate ultrasoft states, leaving
\begin{equation}
S_{\tau_a}(\tau_a) = \frac{1}{N_C}\Tr\bra 0 \overline{Y}_{\bar n}^\dag Y_n^\dag \delta\left(\tau_a - \frac{1}{Q}\int\! d\eta\,\e^{-\abs{\eta}(1-a)}\mathcal{E}_T(\eta)\right)Y_n\overline{Y}_{\bar n}\ket{0}.
\end{equation}
Now consider the properties of this matrix element under a Lorentz boost along the $z$-direction with  a rapidity $\eta'$. Inserting factors of $U(\Lambda(\eta'))^\dag U(\Lambda(\eta')) = 1$,
\begin{equation}
S_{\tau_a}(\tau_a) = \frac{1}{N_C}\Tr\bra 0U^\dag U \overline{Y}_{\bar n}^\dag Y_n^\dag U^\dag U\delta\left(\tau_a - \frac{1}{Q}\int\! d\eta\,\e^{-\abs{\eta}(1-a)}\mathcal{E}_T(\eta)\right)U^\dag U Y_n\overline{Y}_{\bar n}U^\dag U\ket{0}.
\end{equation}
The vacuum $\ket{0}$ is invariant under Lorentz boosts. Wilson lines are also invariant:
\begin{equation}
Y_n(0) = P\exp\left[ ig\int_0^\infty\! ds\, n\cdot A_{us}(ns)\right] \underset{\Lambda(\eta)}{\longrightarrow} P\exp\left[ig\int_0^\infty \!ds\,\alpha n\cdot A_{us}(\alpha ns)\right] = Y_n(0),
\end{equation}
where $\alpha = \e^{-\eta'}$, as $n\rightarrow \alpha n$ and $\bar n\rightarrow \alpha^{-1}\bar n$. (This is also known in SCET as type-III reparametrization invariance \cite{RPI}.) The only change is in the operator $\mathcal{E}_T(\eta)$:
\begin{equation}
\begin{split}
U(\Lambda(\eta'))\mathcal{E}_T(\eta)U(\Lambda(\eta'))^\dag\ket{N(k_i)} &= U(\Lambda(\eta'))\mathcal{E}_T(\eta)\ket{N(\Lambda^{-1} k_i)} \\
&= U(\Lambda(\eta'))\sum_{i\in N}\abs{\vect{k}_i^\perp}\delta(\eta - \eta_i + \eta')\ket{N(\Lambda^{-1} k_i)} \\
&= \sum_{i\in N}\abs{\vect{k}_i^\perp}\delta (\eta+\eta' - \eta_i)\ket{N(k_i)} \\
&=\mathcal{E}_T(\eta+\eta')\ket{N}.
\end{split}
\end{equation}
Thus, we conclude that
\begin{equation}
\label{angshape}
\begin{split}
S_{\tau_a}(\tau_a) &= \frac{1}{N_C}\Tr\bra 0 \overline{Y}_{\bar n}^\dag Y_n^\dag \delta\left(\tau_a - \frac{1}{Q}\int\! d\eta\,\e^{-\abs{\eta}(1-a)}\mathcal{E}_T(\eta)\right)Y_n\overline{Y}_{\bar n}\ket{0} \\
&= \frac{1}{N_C}\Tr\bra 0 \overline{Y}_{\bar n}^\dag Y_n^\dag \delta\left(\tau_a - \frac{1}{Q}\int\! d\eta\,\e^{-\abs{\eta}(1-a)}\mathcal{E}_T(\eta+\eta')\right)Y_n\overline{Y}_{\bar n}\ket{0}.
\end{split}
\end{equation}
This result does not imply that the leading power correction simply shifts the argument of the perturbative event shape distributions, as the delta function (which is just the shape function to leading order in perturbation theory) is a highly nonlinear function of the energy flow operator and sits sandwiched between Wilson lines in the matrix element.  If we neglect correlations
between these operators, however, we derive a delta function for 
the shape function, and reproduce the 
shift in the distribution, Eq.\ (\ref{shift}) \cite{KS99,flow}.  

However, consider the implication of this formula for the first moment of the $\tau_a$ distribution. Taylor expanding\footnote{This expansion is valid in a region of size $\Delta$ near the endpoint, where $\Lqcd\ll\Delta\ll Q$, so that smearing the distribution over this region yields an expansion in powers of $\Lqcd/\Delta$.} the delta function in Eq.~(\ref{angshape}),
\begin{equation}
S_{\tau_a}(\tau_a) = \delta(\tau_a) - \delta'(\tau_a)\frac{1}{Q}\int\! d\eta\, \e^{-\abs{\eta}(1-a)}\frac{1}{N_C}\Tr\bra{0}\overline{Y}_{\bar n}^\dag Y_n^\dag\mathcal{E}_T(\eta+\eta') Y_n\overline{Y}_{\bar n}\ket{0} + \cdots\, .
\end{equation}
{}From this relation we see that,
remarkably,  the first moment is completely independent of the value of the rapidity appearing in the energy flow operator $\mathcal{E}_T(\eta)$, as we are free to choose any value for $\eta'$. 
Thus, we may take the matrix element containing the $\mathcal{E}_T$ operator out of the integral over $\eta$, leaving us to evaluate simply:
\begin{equation}
\int_{-\infty}^\infty \!d\eta\,\e^{-\abs{\eta}(1-a)} = \frac{2}{1-a}.
\end{equation}
For the $C$-parameter distribution, the whole argument would be the same, except that the rapidity integral appearing here would be
\begin{equation}
\int_{-\infty}^\infty \!d\eta\,\frac{3}{\cosh\eta} = 3\pi.
\end{equation}
Thus, the shape functions satisfy
\begin{align}
S_{\tau_a}(\tau_a) &= \delta(\tau_a) - \delta'(\tau_a)\frac{2}{1-a}\mathcal{A} +\cdots\\
S_C(C) &= \delta(C) - \delta'(C)3\pi\mathcal{A} +\cdots,
\end{align} 
and the universal quantity $\mathcal{A}$ is given by
\begin{equation}
\mathcal{A} = \frac{1}{Q}\frac{1}{N_C}\Tr\bra{0}\overline{Y}_{\bar n}^\dag Y_n^\dag \mathcal{E}_T(\eta)Y_n\overline{Y}_{\bar n}\ket{0},
\end{equation}
which is in fact independent of the value of $\eta$ chosen in $\mathcal{E}_T(\eta)$. When convoluted with the perturbative distribution, $S_{\tau_a}(\tau_a)$ and $S_C(C)$ reproduce the universality relations of Eq.~(\ref{avgshift}) for the first moments of the distributions. We have thus established these results without appealing to the one-gluon approximation.

\begin{acknowledgments}
The authors would like to thank the organizers of the TASI 2004 summer school which inspired the questions addressed in this work, and the organizers of the FRIF Workshop for providing the stimulating environment which inspired the answers. CL would like to thank the nuclear theory group at Caltech for its hospitality during the completion of this work. The work of CL was supported in part by the U.S. Department of Energy under grant number DE-FG02-00ER41132. The work of GS was supported in part by the National Science Foundation, grants PHY-0354776 and PHY-0345922.

\end{acknowledgments}

%\begin{thebibliography}{9}   % Use for  1-9  references

\end{document}